\documentclass[aps,prl,showpacs,superscriptaddress,twocolumn]{revtex4-1}

\usepackage{bm}
\usepackage{amsfonts}
\usepackage{subfigure}
\usepackage{amsmath}
\usepackage{amssymb}
\usepackage{amsbsy} 
\usepackage{epsfig}
\usepackage{epstopdf}
\usepackage{latexsym,graphicx,color}
\usepackage{enumerate}
\usepackage{hyperref}
\usepackage{mathtools}

\def\beq{\begin{equation}} 
\def\eeq{\end{equation}}

\begin{document}

\title{Radial Fulde-Ferrell-Larkin-Ovchinnikov state in a population-imbalanced Fermi gas}

\author{Daisuke Inotani}
\email{dinotani@keio.jp}
\affiliation{Department of Physics $\&$ Research and Education Center for Natural Sciences,\\ Keio University,Hiyoshi 4-1-1, Yokohama, Kanagawa 223-8521, Japan}

\author{Shigehiro Yasui}
\email{yasuis@keio.jp}
\affiliation{Department of Physics $\&$ Research and Education Center for Natural Sciences,\\ Keio University,Hiyoshi 4-1-1, Yokohama, Kanagawa 223-8521, Japan}

\author{Takeshi Mizushima}
\email{mizushima@mp.es.osaka-u.ac.jp}
\affiliation{Department of Materials Engineering Science, Osaka University, Toyonaka, Osaka 560-8531, Japan}

\author{Muneto Nitta}
\email{nitta(at)phys-h.keio.ac.jp}
\affiliation{Department of Physics $\&$ Research and Education Center for Natural Sciences,\\ Keio University,Hiyoshi 4-1-1, Yokohama, Kanagawa 223-8521, Japan}
\date{\today}



\begin{abstract}
The possibility of a Fulde-Ferrell-Larkin-Ovchinnikov (FFLO) state in a population imbalanced Fermi gas with a vortex is proposed. Employing the Bogoliubov-de-Gennes formalism we self-consistently determine the superfluid order parameter and the particle number density in the presence of a vortex. We find that as increasing population imbalance, the superfluid order parameter spatially oscillates around the vortex core in the radial direction, indicating that the FFLO state becomes stable. We find that the radial FFLO states cover a wide region of the phase diagram in the weak-coupling regime at $T=0$ 
in contrast to the conventional case without a vortex.
We show that this inhomogeneous superfluidity can be detected as peak structures of the local polarization rate associated with the node structure of the superfluid order parameter. 
Since the vortex in the 3D Fermi gas with population imbalance has been already realized in experiments, our proposal is a promising candidate of the FFLO state in cold atom physics.
\end{abstract}

\maketitle
The Fulde-Ferrell-Larkin-Ovchinnikov (FFLO) states are proposed as inhomogeneous Fermionic superfluid/superconductors with spatial oscillation of the order parameter~\cite{PhysRev.135.A550,larkin:1964zz}.
The possibility of the FFLO states  
has been extensively discussed not only in condensed matter
 physics such as superconductors
 \cite{doi:10.1143/JPSJ.76.051005,PhysRevLett.121.157004,PhysRevLett.119.217002,kasahara,PhysRevLett.100.117002,Mayaffre2014,PhysRevLett.106.137004,Kenzelmann_2017}
 and $^3$He under confinement~\cite{vorontsov,PhysRevB.89.140502,wim16,PhysRevLett.122.085301,shook}   
 but also in high energy physics such as 
 high density QCD~\cite{PhysRevD.63.074016,RevModPhys.76.263,RevModPhys.86.509} and nuclear matter (proton superconductors and neutron superfluids) in a neutron star~\cite{PhysRevC.63.025801,PhysRevC.65.031302} and in a magnetar~\cite{doi:10.7566/JPSCP.20.011006}. 
 The FFLO states 
 have been originally proposed as a ground state of superconductor with a Zeeman energy associated with magnetic field~\cite{PhysRev.135.A550,larkin:1964zz}, but the realization of the FFLO state in electron system is still challenging, because the magnetic field causes orbital effects, which suppress the suprconductivity, in addition to the Zeeman effects. Indeed, in the electron systems there are  few promising candidates for the FFLO state.

Ultracold Fermi gas has been attracted much attentions as an ideal system to realize the FFLO states both experimentally~\cite{Zwierlein492,Partridge503,PhysRevLett.97.190407,PhysRevLett.103.170402,liao2010spin,PhysRevLett.117.235301} and theoretically~\cite{PhysRevLett.94.060404,PhysRevA.73.051603,PhysRevA.75.023614,PhysRevA.74.013614,SHEEHY20071790,PhysRevLett.101.215301,Parish2007,doi:10.1143/JPSJ.76.104006,PhysRevLett.97.120407,PhysRevLett.98.070402,PhysRevLett.98.070403,PhysRevB.76.085120,Parish2007,PhysRevLett.99.250403,PhysRevA.83.023604,Baksmaty_2011}, because one can tune independently the Zeeman effects and the orbital effects.
One of the most promising candidates is a one-dimensional (1D) Fermi gas with a population imbalance~\cite{PhysRevLett.98.070402,PhysRevA.75.063601,PhysRevA.76.043605,PhysRevLett.98.070403,PhysRevB.76.085120,PhysRevLett.99.250403}. In this system, the FFLO state has been predicted to cover a large region of the phase diagram with respect to the interaction strength and population imbalance. Recently, the density profile of population imbalanced 1D Fermi gas was found to qualitatively agree with a theoretical prediction, exhibiting the FFLO state~\cite{liao2010spin,PhysRevLett.117.235301}. However, the evidence of the FFLO state has not been directory detected. Although it has been known that the FFLO state is also favored in two-dimensional (2D) system~\cite{PhysRevA.77.053617}
, it has not been realized yet. 

On the other hand, in three-dimensional (3D) case, the realization of the FFLO state is still more challenging. In this case, it has been predicted that the FFLO states occupy only a narrow region in the phase diagram at zero temperature~\cite{PhysRevLett.94.060404,SHEEHY20071790}, and this region vanishes with increasing the temperature~\cite{Parish2007}, because the phase separation into a non-polarized superfluid and a fully-polarized normal fluid occurs. We note that in the presence of the trapping potential, the spatial oscillation of the superfluid order parameter at the trap edge has been proposed within the Bogoliubov-de-Gennes (BdG) formalism. However, because the amplitude of the oscillation is much smaller than the value of the superfluid order parameter in the bulk, it is difficult to detect. In Refs.~\cite{yanase1,yanase2}, the angular-FFLO state, in which the superfluid order parameter oscillates in the angular direction of a toroidal trap, has been discussed. See also Ref.~\cite{Yoshii:2014fwa} for an FFLO state in a superconducting ring. Furthermore the FFLO state stabilized by an optical lattice has been proposed~\cite{Koponen_2008}. However, in both cases, any direct evidences of the FFLO state have not been observed, so far.

In this Letter, we theoretically propose an experimentally accessible rote to reach the FFLO state in 3D system.
In our idea, we consider a quantum vortex in the 3D superfluid Fermi gas with a population imbalance. In contrast to the case with no vortices, where the excess atoms gather at the trap edge, in the presence of a vortex, they can localize near the vortex core. As a result, the polarized Fermi gas is realized around the vortex core and the FFLO state appears in the wide region of the phase diagram with respect to the interaction strength and population imbalance at zero temperature. We emphasize that this situation should have been already experimentally realized~\cite{Zwierlein492,Zwierlein2006}, although the observation of the FFLO state has not been reported. Thus only a more precise measurement is needed to clearly detect the FFLO state. In this Letter, we take $\hbar = k_{\rm B}=1$.

To clarify our idea we investigate a singly isolated quantum vortex in the two-component Fermi gas with population imbalance within the BdG formalism~\cite{PhysRevB.43.7609,PhysRevLett.97.180407,PhysRevA.77.063617}, starting from the Hamiltonian
\begin{align}
H_{\rm {BdG}}&=\sum_{\sigma =\uparrow,\downarrow} \int d {\bm r} \psi_{\sigma}^{\dagger}\left( {\bm r} \right)
\left(
-\frac{\nabla^2}{2m}-\mu_\sigma
\right)
\psi_{\sigma}\left( {\bm r} \right)
\nonumber
\\
&+\int d {\bm r} \left(
\Delta \left( {\bm r} \right)
\psi_{\uparrow}^{\dagger}\left( {\bm r} \right)
\psi_{\downarrow}^{\dagger}\left( {\bm r} \right)
+h.c.
\right)
\nonumber
\\
&-
U_s\sum_{\sigma =\uparrow,\downarrow} \int d {\bm r}
n_{-\sigma} \left({\bm r} \right) 
\psi_{\sigma}^{\dagger}\left( {\bm r} \right)
\psi_{\sigma}\left( {\bm r} \right)
.
\label{eqHamiltonian}
\end{align}
Here $\psi_\sigma(\bm {r})$ is the field operator of a Fermi atom with pseudospin $\sigma = \uparrow, \downarrow$ and the atomic mass $m$. $\mu_\sigma$ is the chemical potential of the $\sigma$ component. The population imbalance is included in the difference between $\mu_\uparrow$ and $\mu_\downarrow$. The second and third terms describe the contribution from the superfluid order parameter $\Delta({\bm r}) = -U_s \bigl\langle\psi_{\downarrow} ( {\bm r} )\psi_{\uparrow} ( {\bm r} ) \bigr\rangle$ and the Hartree potential  $-U_sn_{-\sigma}(r)=-U_s\bigl\langle\psi^{\dagger}_{-\sigma} \left( {\bm r} \right)\psi_{-\sigma} \left( {\bm r} \right) \bigr\rangle$, respectively, where $n_{\sigma}({\bm r})$ is the number density of the $\sigma$ component.

We consider a single vortex along the $z$ axis with the winding number $w=1$ at $\rho = 0$ in the cylindrical coordinates ${\bm {r}}= \left( \rho, \theta, z\right)$. 
In this cylindrically symmetric situation, we can write the superfluid order parameter and particle number density as $\Delta\left({\bm r}\right)=\Delta\left( \rho \right)e^{i\theta}$ and $n_\sigma\left({\bm r} \right)=n_\sigma\left(\rho \right)$, respectively. In this Letter, we consider the FFLO state with a spatial oscillation of $\Delta \left( \rho \right)$ along the radial direction.

The mean fields, {\it i.e.} $\Delta\left(\rho\right)$ and $n_\sigma\left(\rho\right)$, as well as the chemical potential $\mu_\sigma$, are determined by self-consistently solving the gap equation and the particle number equations for a given interaction strength and population imbalance $P=(N_\uparrow-N_\downarrow)/(N_\uparrow+N_\downarrow)$, where $N_{\sigma} = \int d{\bm r} n_{\sigma}(\bm{r})$ is the total atomic number of the $\sigma$ component. This procedure can be achieved  by conventional diagonalization, {\it i.e.}, the Bogoliubov transformation for the cylindrical symmetric system \footnote{See Supplemental Materials for the details of calculations.
}.
In addition to  $\Delta\left(\rho\right)$ and $n_\sigma\left(\rho\right)$, we calculate the local density of states (LDOS) given by
\begin{align}
{\mathcal N}_\uparrow\left( \omega, \rho \right)
&=-\frac{1}{\pi}{\rm Im}
G_{11} \left(\bm {r},\bm{r} , i\omega_n \to \omega+i\epsilon \right)
\\
{\mathcal N}_\downarrow\left( \omega, \rho \right)
&=\frac{1}{\pi}{\rm Im}
G_{22} \left(\bm {r},\bm{r} , i\omega_n \to \omega+i\epsilon \right),
\end{align}
where $\omega_n= (2n+1)\pi T$ ($n\in \mathbb{Z}$) is the Matsubara frequency at temperature $T$, and $\epsilon$ is an infinitesimally small parameter. Here 
\begin{align}
\hat{G}\left(\bm {r},\bm{r}' , i\omega_n \right)
=-\int_0^{\beta} e^{i\omega_n \tau} 
\left\langle
T_{\tau} \left\{
\Psi
\left( {\bm r} ,\tau \right)
,
\Psi^{\dagger}
\left( {\bm r}' ,0 \right)
\right\}
\right\rangle
\end{align}
is a 2$\times$2 single-particle Green's function with the two-component Nambu-Gor'kov field operator $\Psi
\left( {\bm r} ,\tau \right)
=\bigl(
\psi_\uparrow
\left( {\bm r} ,\tau 
\right)
~\psi_\downarrow^{\dagger}
\left( {\bm r} ,\tau 
\right)
\bigr)
$. Finally, we summarize the setup of the numerical calculations. We take $Rk_{\rm F}=50$ and $L_zk_{\rm F}=20$ for the system size of the $\rho$ and $z$ directions, respectively, where $k_{\rm F}$ is the Fermi momentum. We take the cutoff energy $E_c=9\varepsilon_{\rm F}$ with the Fermi energy $\varepsilon_{\rm F}=k_{\rm F}^2/(2m)$. We fix $T=0$.
\begin{figure}
\centerline{\includegraphics[width=7cm]{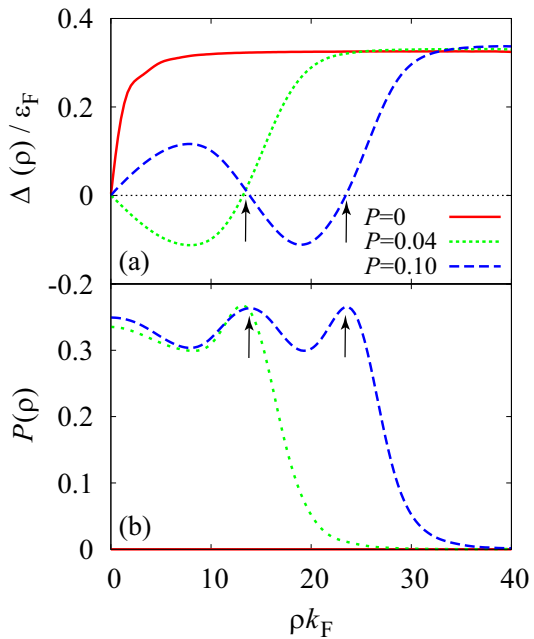}}
\caption{Calculated (a) superfluid order parameter and (b) local population imbalance $P(\rho)=(n_\uparrow(\rho) - n_\downarrow(\rho))/(n_\uparrow(\rho) + n_\downarrow(\rho))$
, as a function of $\rho$. The solid line shows the results with $P=0$. The dotted and dashed line are corresponding the case with $N=1$ and $N=2$, respectively, where $N$ is the number of the node structure. In this figure we take $(k_{\rm F}a_s)^{-1}=-0.5$ The arrows denote the node structure in the case with $N=2$ (dashed line). 
}
\label{figdelta}
\end{figure}

In Fig. \ref{figdelta}, we show the self-consistent solutions of $\Delta (\rho)$ in the weak-coupling regime with $(k_{\rm F}a_s)^{-1}=-0.5$, where $a_s$ is the $s$-wave scattering length~\footnote{
See the Supplemental material for the definition of $a_s$.}. 
In the absence of the population imbalance ($P=0$), the ordinary vortex is obtained. As $P$ increases, we find that $\Delta(\rho)$ spatially oscillates around the vortex core and approaches the value in the bulk away from the vortex core, that indicates the FFLO state locally realizes near the vortex core. Further increasing $P$, the number $N$ of nodes (where $\Delta (\rho)=0$) increases. The dotted and dashed lines in Fig.~\ref{figdelta} (a) are corresponding to the $N=1$ and $N=2$ cases, respectively.

This dependence of $\Delta(\rho)$ on $P$ can be understood as follows. In the presence of the population imbalance, the excess atoms gather into the region where the superfluid order parameter is small, because the excess atoms feel the superfluid order parameter as a potential. The sign change of the superfluid order parameter at vortices and FFLO nodal planes, leads to the formation of low-lying quasiparticle states. Bogoliubov quasiparticle states in the vortex core are discretized to the Caroli-de Gennes-Matricon (CdGM) states with level spacing $\sim \Delta^2_0/\varepsilon_{\rm F}$, where $\Delta_0$ is the bulk value of the superfluid order parameter~\cite{CAROLI1964307,PhysRevLett.80.2921}, while the FFLO nodal planes are accompanied by mid-gap Andreev bound states~\cite{PhysRevB.30.122,mizushima18,doi:10.1143/JPSJ.65.246}. When the population imbalance is small, the excess atoms are accumulated by the CdGM states and thus localize around the vortex core. However, as increasing the number of excess atoms, the vortex size also increases to contain more atoms, leading to the increase of energy of the vortex. Eventually, it becomes energetically favorable to make a node structure, which is accompanied by mid-gap Andreev bound states and can accumulate the excess atoms. Hence, the existence of a vortex line can become a trigger for realizing the FFLO state. Indeed, as shown in Fig.~\ref{figdelta} (b), the local polarization rate defined by $P(\rho)=(n_\uparrow(\rho)-n_\downarrow(\rho))/(n_\uparrow(\rho)+n_\downarrow(\rho))$ has peak structures around the nodes ($\rho k_{\rm F} \simeq 13$, $24$ for the dashed line in Fig. \ref{figdelta} (b)), which can be measured as an evidence of our proposal. 

We also emphasize that the amplitude of the oscillation of $\Delta(\rho)$ is comparable to the bulk value of the superfluid order parameter. This is in contrast to the trapped case, where while the similar oscillation is predicted at the trap edge, the amplitude is much smaller than the value of $\Delta({\bm r})$ at the trap center~\cite{doi:10.1143/JPSJ.76.104006}. The resultant local polarization cannot possess pronounced peak structures at the nodal planes. Thus, the FFLO state proposed in this work is more promising to experimentally detect.  
\begin{figure}
\centerline{\includegraphics[width=8cm]{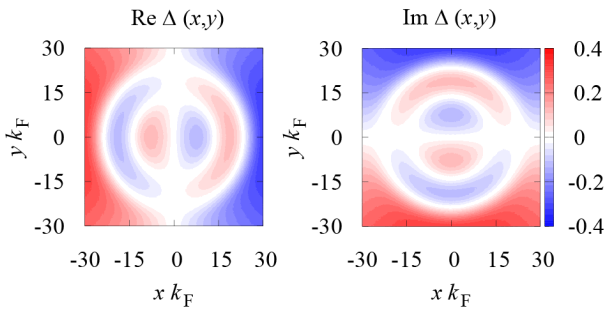}}
\caption{Spatial structure of the superfluid order parameter $\Delta({\bm r})=\Delta(\rho)e^{i\theta}$ in the $x$-$y$ plane. The parameters are taken to be the same as those in the $N=2$ case in Fig. \ref{figdelta}.}
\label{figdelta_spatial}
\end{figure}

The spatial structure of the superfluid order parameter $\Delta({\bm r})=\Delta(\rho)e^{i\theta}$ is shown in Fig.~\ref{figdelta_spatial}. We find the clear oscillation of $\Delta({\bm r})$ in the radial direction $\rho$. In addition to these nodes, the real (imaginary) part of $\Delta({\bm r})$ vanishes along $y$ ($x$) axis. This is simply because of the phase factor $e^{i\theta}$ associated with the vortex. 

\begin{figure}
\centerline{\includegraphics[width=9cm]{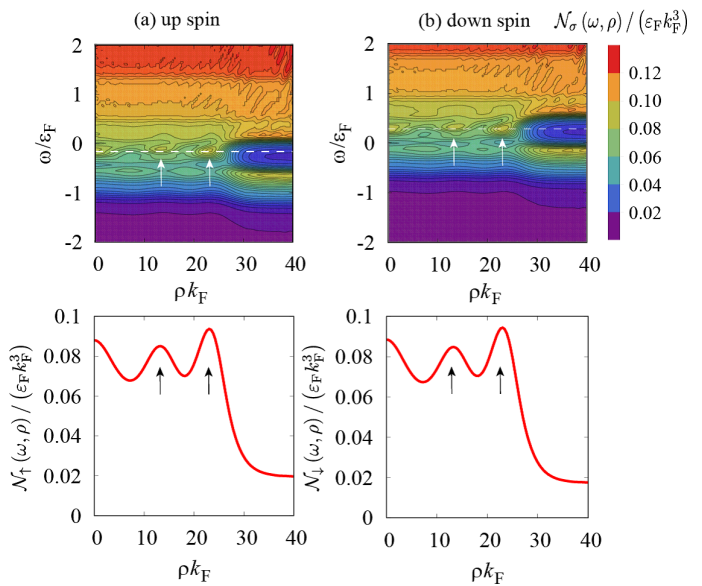}}
\caption{Calculated LDOS of (a) $\uparrow$ and (b) $\downarrow$ component (upper panels). The parameters are taken to be the same as those in the $N=2$ case in Fig. \ref{figdelta}. Values of LDOS along the dashed lines in upper panels are also shown (lower panels). We use $\omega=-0.16\varepsilon_{\rm F}$ for $\uparrow$ spin and at $\omega = 0.24\varepsilon_{\rm F}$ for $\downarrow$ spin. The arrows indicate the mid-gap Andreev states around the nodal planes of the FFLO state.}
\label{figldos}
\end{figure}

The mid-gap Andreev bound states and the CdGM states, which are associated with the nodal planes of the FFLO state and the vortex, respectively, can be detected by an observation of LDOS ${\mathcal N}_\sigma (\omega,\rho)$. Figure \ref{figldos} shows the calculated LDOS with the same parameters as in the case with $N=2$ in Fig.~\ref{figdelta} (dashed lines). While in the bulk region the clear gap structure opens in LDOS, in the region where the superfluid order parameter spatially oscillates ($\rho k_{\rm F} \lesssim 30$), LDOS has a finite value with an energy inside the superfluid gap. To clearly see this, in the lower panels in Fig,~\ref{figldos}, we show the $\rho$ dependence of LDOS with a fixed energy ($\omega=-0.16\varepsilon_{\rm F}$ for $\uparrow$ spin and $\omega=0.28\varepsilon_{\rm F}$ for $\downarrow$ spin). In each panel, we find three peak structures. The peak around the vortex core $\rho \simeq 0$ corresponds to the CdGM states, and the others correspond to the mid-gap Andreev bound states. Thus, the disappearance of the gap structure in LDOS except around the vortex core can be an evidence of the realization of the FFLO state. Since the occupied LDOS can be experimentally observed by using a local photoemission spectroscopy~\cite{PhysRevLett.114.075301}, the characteristic structures in LDOS of the $\uparrow$ component are accessible.

\begin{figure}
\centerline{\includegraphics[width=8cm]{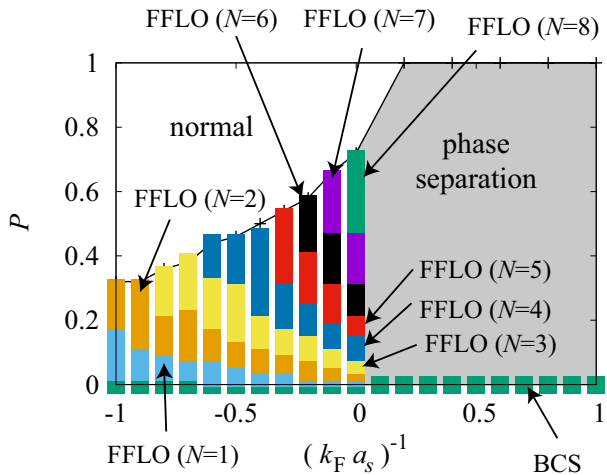}}
\caption{Phase diagram of the population imbalanced Fermi gas with a vortex. $N$ in the FFLO state denotes the number of the node structure. In the shaded area in the strong-coupling regime, the phase separation into the non-polarized superfluid and the fully-polarized normal fluid occurs. The BCS state without spatial oscillation of the superfluid order parameter is obtained only in the absence of the population imbalance $P=0$ within our calculation.} 
\label{figphase}
\end{figure}

Finally, we show the phase diagram with respect to $(k_{\rm F} a_s)^{-1}$ and $P$ at $T=0$ in Fig. \ref{figphase}. We find that the FFLO state covers the wide region of the phase diagram in the weak-coupling regime $(k_{\rm F} a_s)^{-1} \le 0$ in contrast to the BCS superfluid phase {\it without} the spatial oscillation of $\Delta\left({\bm r} \right)$, which is realized only in the case with small population imbalance. 
 We also mention that in the strong-coupling regime where $(k_{\rm F}a_s)^{-1} > 0$, the phase separation into the spin-balanced superfluid region and the fully polarized normal fluid region occurs, 
which also happens in the case without a vortex.  
 Thus, the FFLO oscillation cannot be realized. This result is reasonable. When we consider the strong-coupling limit, the most of the Fermi atoms form Cooper pairs except the excess atoms. Thus, the Fermionic nature vanishes in this limit, except a small Fermi surface formed by the excess atoms. On the other hand, the FFLO state is stabilized by the mismatch of the size of the Fermi surface between the $\uparrow$ and $\downarrow$ components. However, in the strong-coupling limit the Fermi surface of the $\downarrow$ component vanishes. Thus, the FFLO state is realized only in the weak-coupling regime.

To summarize, we have proposed a new route to reach the FFLO superfluid in 3D Fermi gas. We have considered the population imbalanced Fermi gas with a vortex. Applying the BdG formalism to this system, we have shown that the spatial oscillation of the superfluid order parameter appears near the vortex core and the number of the node structure increases as the population imbalance increases. We have also found that the FFLO nature can be seen as peak structures in the local polarization rate, as well as vanishing gap structure in the LDOS. We have shown that the FFLO states cover a wide region of the phase diagram in the weak-coupling regime at zero temperature in contrast to the conventional case without a vortex.

This work is supported by the Ministry of Education, Culture, Sports, Science (MEXT)-Supported Program for the Strategic Research Foundation at Private Universities ``Topological Science'' (Grant No. S1511006). This work is also supported in part by Japan Society for the Promotion of Science (JSPS) Grants-in-Aid for Scientific Research (KAKENHI Grant No. 17K05435 (S. Y.), No. JP16K05448 (T. M.), No. 16H03984 (M. N.), and No. 18H01217 (M. N.)), and also by MEXT KAKENHI Grant-in-Aid for Scientific Research on Innovative Areas ``Topological Materials Science,'' Grant No. 15H05855 (T. M. and M. N.).

\bibliography{fflo}


\clearpage

\renewcommand{\thesection}{S\arabic{section}}
\renewcommand{\theequation}{S\arabic{equation}}
\setcounter{equation}{0}
\renewcommand{\thefigure}{S\arabic{figure}}
\setcounter{figure}{0}

\onecolumngrid
\appendix
\begin{center}
\large{Supplemental Material for}\\
\textbf{``Radial Fulde-Ferrell-Larkin-Ovchinnikov state in a population-imbalanced Fermi gas"}
\end{center}

\section{Diagonalization of the BdG Hamiltonian in cylindrical system}

In this section, we summarize the procedure of the diagonalization of the BdG Hamiltonian in Eq.~\eqref{eqHamiltonian} under the cylindrical symmetry. For this purpose, it is useful to expand $\psi_\sigma \left( {\bm r} \right)$ with respect to a set of eigenfunctions of the kinetic energy term in the cylindrical coordinate as
\begin{align}
\psi_{\sigma} \left( {\bm r} \right) = 
\sum_{j=1}^{\infty} \sum_{l=-\infty}^{\infty} \sum_{k_z}
c_{j,\sigma}^{l,k_z} f_{j,l,k_z}\left({\bm r}\right) 
\end{align}
where
\begin{align}
f_{j,l,k_z}\left({\bm r}\right)
=
\phi_{j,l}\left( \rho \right) e^{il\theta} \frac{e^{ik_z z}}{\sqrt{2\pi L}}. 
\end{align}
Here $L$ is the height to the $z$ direction of the system ($0 \le z \le L $) and the normalized radial wave function $\phi_{j,l} \left(\rho \right)$ is given by
\begin{align}
\phi_{j,l}\left( \rho \right) = \frac{\sqrt{2}}{RJ_{l+1}\left( \alpha_{j,l} \right)} J_{l}\left( \alpha_{j,l} \frac{\rho}{R} \right)
\end{align}
where $J_l$ is Bessel function, $\alpha_{j,l}$ is $j$th zero of $J_{l}$, and $R$ is the system radius ($0 \le \rho \le R$). In this basis, the BdG Hamiltonian in Eq.~\eqref{eqHamiltonian} can be written as
\begin{align}
H_{\rm {BdG}} =
\sum_{l,k_z}\sum_{j,j'}
{\Phi_{j}^{l,k_z}}^\dagger
h_{j,j'}^{l,k_z}
\Phi_{j'}^{l,k_z}.
\end{align}
Here, we have introduced the Nambu-Gor'kov field operator in the cylindrical coordinate as
\begin{align}
\Phi_j^{j,k_z}
&=
\left(
\begin{array}{c}
c_{j,\uparrow}^{l,k_z}
\\
{c_{j,\downarrow}^{-l-1,-k_z}}^\dagger
\end{array}
\right)
\\
\Phi^{\dagger} \left({\bm r}\right)&=
\left(
\begin{array}{cc}
{c_{j,\uparrow}^{l,k_z}}^\dagger
& 
c_{j,\downarrow}^{-l-1,-k_z}
\end{array}
\right)
\end{align}
and the matrix $h_{j,j'}^{l,k_z}$ is given by
\begin{align}
h_{j,j'}^{l,k_z} =
\left(
\begin{array}{cc}
\xi^{\uparrow}_{j,l,k_z}\delta_{j,j'}+F_{j,j'}^{l,\uparrow}
& 
\Delta_{j,j'}^{l} 
\\
\Delta_{j,j'}^{l} 
& 
-\xi^{\downarrow}_{j,l+1,k_z}\delta_{j,j'}-F_{j,j'}^{l+1,\downarrow}
\end{array}
\right)
\end{align}
with the superfluid order parameter
\begin{align}
\Delta_{j,j'}^{l} &= \int_0^{R} \rho d \rho \phi_{j,l} \left( \rho \right) \Delta \left( \rho \right) \phi_{j',l+1} \left( \rho \right)
\end{align}
and the Hatree potential
\begin{align}
F_{j,j'}^{l,\sigma} &= -U_s\int_0^{R} \rho d \rho \phi_{j,l} \left( \rho \right) n_{-\sigma} \left( \rho \right) \phi_{j',l} \left( \rho \right).
\end{align}
The Hamiltonian can be diagonalized by the Bogoliubov-Valatine transformation
\begin{align}
\gamma_{j,\sigma}^{l,k_z}
=\sum_{j',\sigma'}
\left(
W^{-1}
\right)^{l,k_z}
_{\{j,\sigma \} ,\{ j',\sigma' \}}
\Phi_{j',\sigma'}^{l,k_z}
\end{align}
with an orthogonal matrix $\hat{W}$ as
\begin{align}
H_{\rm{BdG}}= \sum_{\sigma} \sum_{j,l,k_z} 
E^{l,k_z}_{j,\sigma}
\left(\gamma_{j,\sigma}^{l,k_z}\right)^{\dagger}
\gamma_{j,\sigma}^{l,k_z}
\end{align}
where $E^{l,k_z}_{j,\sigma}$ is the eigenvalues of the Hamiltonian. We note that the matrix in the original BdG Hamiltonian in Eq.~\eqref{eqHamiltonian}  is diagonal in terms of $l$ and $k_z$. Thus, it is sufficient to numerically solve the eigenvalue equation with $l$ and $k_z$ fixed.
\par
Using the set of eigenfunction $W$ and eigenvalues $E$, the self-consistent equations for the superfluid order parameter and the particle number density can be obtained as
\begin{align}
\Delta\left({\bm r} \right) 
&= 
-\frac{U_se^{-i\theta}}{2\pi L}
\sum_{l,k_z}
\sum_{j,j'}
\phi_{j,l+1}\left(\rho\right)
\phi_{j',l}\left(\rho\right)
d_{j,j'}^{l,k_z},
\\
n_\uparrow \left({\bm r} \right) 
&=
\frac{1}{2\pi L}
\sum_{l,k_z}
\sum_{j,j'}
\phi_{j,l}\left(\rho\right)
\phi_{j',l}\left(\rho\right)
\eta^{\uparrow}_{j,j'},
\\
n_\downarrow \left({\bm r} \right) 
&=
\frac{1}{2\pi L}
\sum_{l,k_z}
\sum_{j,j'}
\phi_{j,l+1}\left(\rho\right)
\phi_{j',l+1}\left(\rho\right)
\eta^{\downarrow}_{j,j'},
\end{align}
respectively. Here, we have defined 
\begin{align}
d_{j,j'}^{l,k_z}
&= \sum_{i,\sigma} 
W^{l,k_z}
_{ \{ j,\downarrow \}
,\{ i,\sigma \} }
W^{l,k_z}
_{ \{ j',\uparrow \}
,\{ i,\sigma \} }
n_{\rm F}\left(
E^{l,k_z}_{i,\sigma}
\right)
\\
\eta^{\uparrow}_{j,j'}
&=\sum_{i,\sigma} 
W^{l,k_z}
_{ \{ j,\uparrow \}
,\{ i,\sigma \} }
W^{l,k_z}
_{ \{ j',\uparrow \}
,\{ i,\sigma \} }
n_{\rm F}\left(
E^{l,k_z}_{i,\sigma}
\right)
\\
\eta^{\downarrow}_{j,j'}
&=\sum_{i,\sigma} 
W^{l,k_z}
_{ \{ j,\downarrow \}
,\{ i,\sigma \} }
W^{l,k_z}
_{ \{ j',\downarrow \}
,\{ i,\sigma \} }
\left( 1- n_{\rm F}\left(
E^{l,k_z}_{i,\sigma}
\right)
\right)
\end{align}
To avoid the well-known ultra-violet divergence, we need to introduce a cutoff energy $E_{\rm c}$ in the gap equation. We also note that the interaction strength is conveniently measured by the $s$-wave scattering length $a_s$ in cold atom physics. In our formalism, $a_s$ is related to the coupling constant $U$ and the cutoff energy $E_{\rm c}$ as
\begin{align}
\frac{1}{k_{\rm F} a_s}&=
-8\pi
\frac{\varepsilon_{\rm F}}{U_s k_{\rm F}^3}
+\frac{2}{\pi}
\sqrt{\frac{E_c}{\varepsilon_{\rm F}}}.
\end{align}

\end{document}